\documentclass[10pt,a4paper]{article}
\usepackage{graphicx}

\begin{document}
\title{Connecting the underlying event with jet properties in pp collisions at $\sqrt{s}$ = 7 TeV with the ALICE experiment}

\author{Hermes Le\'on Vargas (for the ALICE collaboration) \\
\small Institut f\"ur Kernphysik, Goethe-Universit\"at Frankfurt am Main \\
\small Max-von-Laue-Str. 1, Frankfurt am Main, 60438, Germany \\
\small hleon@ikf.uni-frankfurt.de
}

\maketitle


\begin{abstract}
A preliminary study of the fragmentation properties of charged particle jets as a function of the Underlying Multiplicity (UM) is presented. The UM is defined such that it can be related to the global soft event characteristics by excluding the contribution due to jet fragmentation. The measurement of jet properties as a function of the UM might be connected to the impact parameter dependence of the transverse nucleon structure as described via Generalized Parton Distributions. The results from the studies are compared to Monte Carlo (MC) models.
\end{abstract}

\section{Introduction}
The standard procedure to study the Underlying Event (UE) characteristics starts with defining regions of the azimuthal detector acceptance based on the direction of the leading charged jet \cite{CDF} or the leading track \cite{UE-ALICE} in an event. These regions, each of them covering an equal angle of $|\Delta\phi| = 2\pi/3$ are called Towards, Transverse and Away. The Toward and Away regions are expected to contain the tracks that are produced by the fragmentation of the partons that took part in the hard scattering while the Transverse region is expected to collect tracks from the UE. ALICE \cite{ALICE} has characterized the UE in pp collisions at $\sqrt{s}$ = 0.9 and 7 TeV in \cite{UE-ALICE} by measuring the charged particle density, summed $p_{\mathrm{t}}$ and azimuthal correlations (between the leading track and the remaining tracks from the event) in the above mentioned regions. This study presents an estimator for the multiplicity of the underlying event of proton-proton collisions and then uses it to characterize the properties of charged jets in different ranges of the underlying activity. Thus, this study allows to characterize the interplay between the low momentum transfers related to the underlying activity and the hard partonic interactions that produce jets.

\section{Data analysis}
The experimental results presented in this work are based on a sample of 172 million minimum bias pp collisions at $\sqrt{s}$ = 7 TeV. Charged tracks were reconstructed using the Inner Tracking System (ITS) and the Time Projection Chamber (TPC). The combined information from the ITS and the TPC provide a uniform acceptance of tracks as well as good momentum resolution. Details about the track selection used in the charged jet analysis in ALICE can be found in \cite{ALICEJets}. The jet activity is identified using a jet finder algorithm. For this work we use the SISCone jet finder \cite{SISCone} which is a seedless infrared-safe cone algorithm. The jet finding is performed using charged particles with $p_{\mathrm{t}} >$ 1 GeV/$c$ and a cone radius of $R_{\mathrm{Jet}}$=0.4 as input parameters for the algorithm. The minimum jet transverse momentum was set to 5 GeV/$c$.

\section{Underlying Multiplicity}
The UE definition is that it comprises all the activity in the event but that related to the hardest scattering in a pp collision. In this work we propose to use a jet finding algorithm to identify the orientation in the $\eta-\phi$ space of the products of the jet fragmentation and then characterize the underlying activity in the regions where the jet finder had not identified jet fragmentation activity. We developed a multiplicity estimator to characterize the UM following a suggestion by Frankfurt, Strikman and Weiss who discuss the transverse nucleon structure in \cite{TransverseNucleon}. In that paper it is suggested to associate the multiplicity to the impact parameter of the pp collision. Therefore we propose in this work to use a multiplicity estimator, that is not biased by the multiplicity created by the jet fragmentation, to access experimentally the centrality of the pp collision. This provides also a test of the implementation of pp centrality in modern MC generators, which contain simplifications such as the assumption of factorization of the parton $x$ and the impact parameter as well as the lack of information of the flavor and $x$ dependence of the proton's transverse composition as described in \cite{MonteCarlos}.

\begin{figure}[!htb]
\begin{center}
\includegraphics[width=25pc]{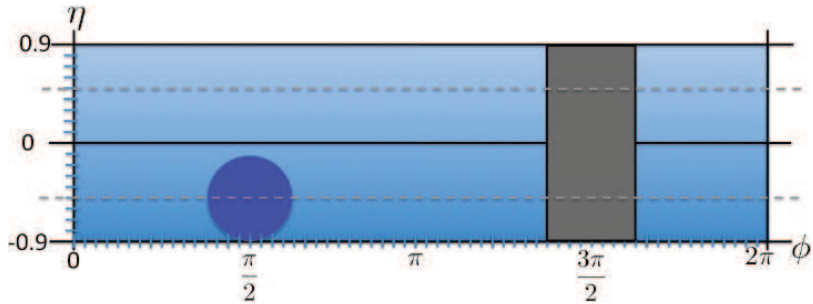}
\end{center}
\caption{\label{fig:figTPCAcc}Definition of soft event multiplicity in jet events. The drawing represents the acceptance of the ALICE TPC in the $\eta$-$\phi$ space. The dark blue (colors in the online version) circle shows the jet area covered by the jet cone  with radius $R_{\mathrm{Jet}}$. The grey rectangular area is the di-jet area. The width of the excluded di-jet area in the $\phi$ direction is 2$R_{\mathrm{Jet}}$ and covers the whole acceptance in the $\eta$ direction. The dotted lines represent the maximum limits for the $\eta$ coordinate of the accepted jet axis, so the complete jet area falls inside the TPC acceptance.}
\end{figure}

\begin{figure}[!htb]
\begin{center}
\includegraphics[width=17pc]{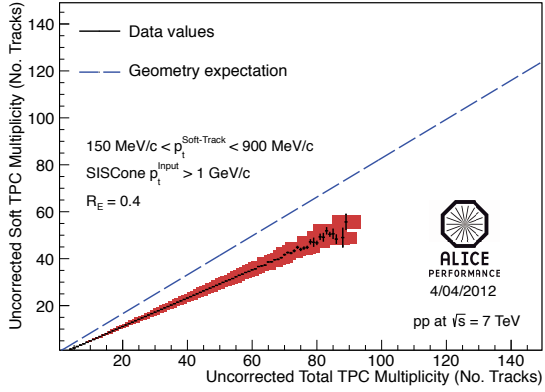}
\end{center}
\caption{\label{fig:figTPCsoftTot} Soft TPC multiplicity as a function of the total TPC multiplicity. The figure shows the experimental measurements as well as the expectation values from a pure geometrical exclusion.}
\end{figure}

The UM estimation is based on the separation of the event phase space into two regions: one that contains jet activity and the complementary region. The UM is calculated using the charged tracks measured within the TPC acceptance ($|\eta| <$ 0.9 and $0 < \phi \leq 2\pi$). The first classification of the tracks from the events is done by their transverse momentum. The $p_{\mathrm{t}}$ cutoff of the jet finder (1 GeV/$c$) has been chosen to allow for a gap between this cutoff and the transverse momentum of the tracks that are used in the multiplicity estimation. Therefore we define as the total TPC multiplicity those tracks that have 150 MeV/$c$ $< p_{\mathrm{t}}^{\mathrm{Soft-Track}} <$ 900 MeV/$c$. The second requirement to select tracks for the UM comes from the geometry of the event, this selection can be easily explained using figure \ref{fig:figTPCAcc}. The figure shows the $\eta$ and $\phi$ acceptance of the ALICE TPC. To calculate the soft TPC multiplicity, that corresponds to the UM measured in the ALICE central barrel, the tracks that fall into two areas of the acceptance are not taken into account in the multiplicity estimation\footnote{The analysis is restricted to events where there is only one reconstructed jet in the whole TPC acceptance. This is done in order to increase the statistical significance of the multiplicity estimator avoiding the subtraction of larger areas. The number of jet events after making this restriction is reduced by 10\% from the total number of jet events in the sample.}. The first area is shown in figure \ref{fig:figTPCAcc} as a circle that corresponds to the jet cone with axis coordinates $\eta^{\mathrm{Jet}}$=0.5 and $\phi^{\mathrm{Jet}}$=$\pi$/2 in this example. The second area that is excluded is the di-jet area\footnote{Even though the analysis is restricted to events where there is only one reconstructed jet the di-jet area is considered in order to make sure that even if the di-jet was not identified by the jet finder the multiplicity estimation is not biased.}, centred at $\phi^{\mathrm{Exc.}}$=$\phi^{\mathrm{Jet}}$+$\pi$, having a width in the $\eta$ axis of $2R_{\mathrm{Jet}}$ and covering the whole acceptance in the $\phi$ direction.
Figure \ref{fig:figTPCsoftTot} shows the distribution of the mean values of the soft TPC multiplicity, shown with black markers, as a function of the total TPC multiplicity defined above. The figure shows as well, using a dashed line, the values of the soft TPC multiplicity that would be expected in the case that the reduction in the multiplicity is caused only by a geometrical exclusion\footnote{The geometrical exclusion is obtained by making the assumption that the tracks are isotropically distributed over the ALICE central barrel acceptance.}. The experimental result shows that the soft TPC multiplicity is a fraction of the geometry expectation and increases linearly as a function of the total TPC multiplicity.

\section{The \textit{NT90} jet observable}
\begin{figure}[htb]
\begin{center}
\includegraphics[width=17pc]{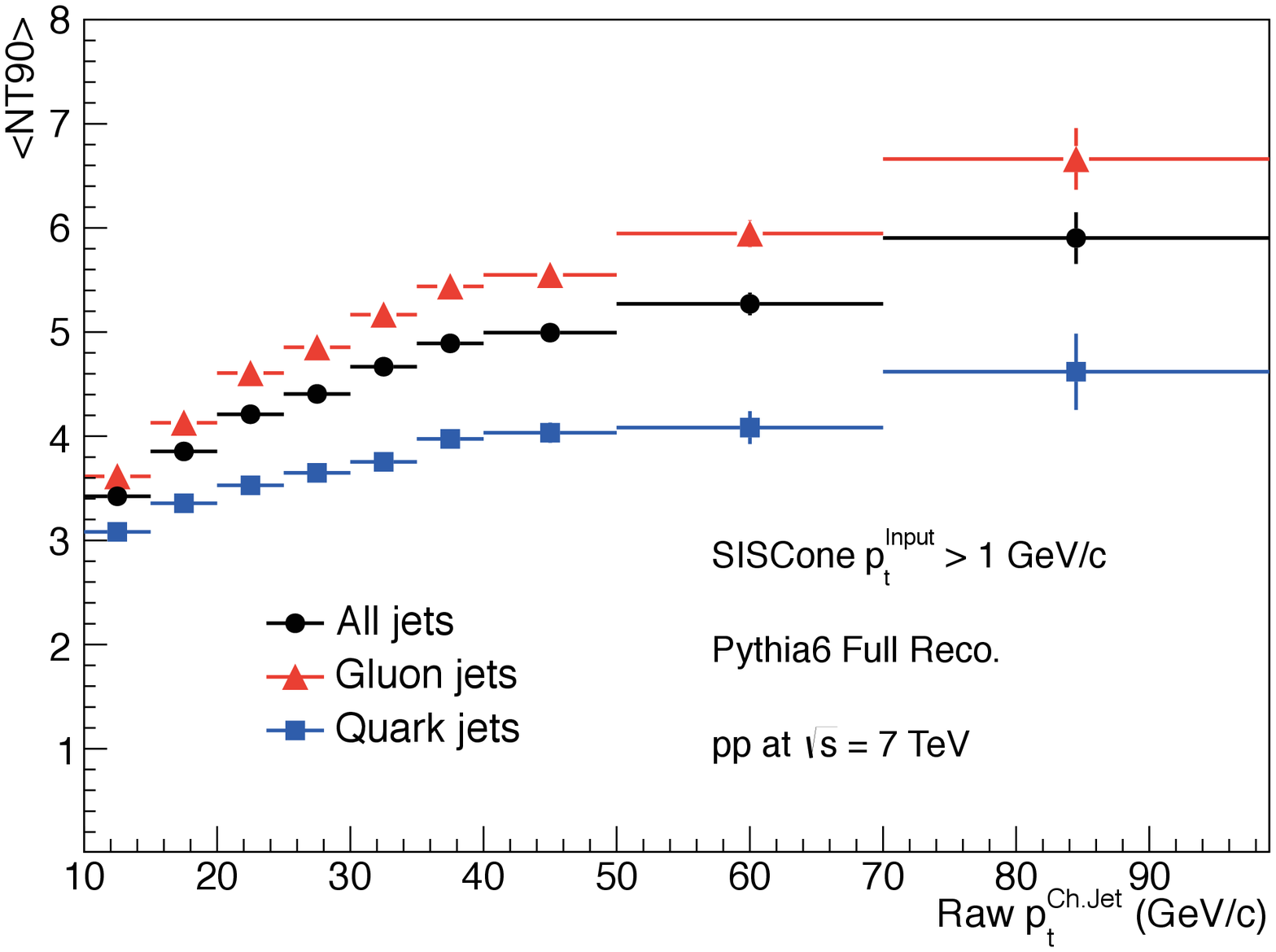}
\end{center}
\caption{\label{fig:figNT90QG} Mean values of \textit{NT90} as a function of the $p_{\mathrm{t}}^{\mathrm{Ch.Jet}}$ for a Pythia6 simulation of pp collisions at $\sqrt{s}$ = 7 TeV. The triangles show the distribution obtained for gluon initiated jets while the squares show the distributions for quark initiated jets. The black circle distribution shows the results obtained for the inclusive jet sample.}
\end{figure}

The jet observable that is used in this work to characterize the fragmentation properties of the charged jets is called \textit{NT90}. \textit{NT90} was proposed in \cite{NT90} as an observable that can be used to discriminate jets originating from quark or gluon fragmentation. \textit{NT90} is constructed by selecting the tracks that are contained in the jet cone and that fulfil the same $p_{\mathrm{t}}$ cutoff used during the jet finding. This set of tracks is then sorted according to their transverse momentum going from the highest to the lowest in magnitude. Then one adds the $p_{\mathrm{t}}$ of the sorted tracks until 90\% of the total charged jet transverse momentum ($p_{\mathrm{t}}^{\mathrm{Ch.Jet}}$) is recovered. \textit{NT90} is the minimum number of tracks necessary to recover 90\% of the total $p_{\mathrm{t}}^{\mathrm{Ch.Jet}}$. Figure \ref{fig:figNT90QG} shows the mean values of \textit{NT90} as a function of the $p_{\mathrm{t}}^{\mathrm{Ch.Jet}}$ obtained with Pythia6 minimum bias events that include the detector simulation of ALICE. The figure shows the different distributions obtained for quark and gluon initiated jets separately as well as for the inclusive jet sample. One can notice that the mean value of \textit{NT90} at a fixed value of the $p_{\mathrm{t}}^{\mathrm{Ch.Jet}}$ is larger for gluon initiated jets than for quark initiated jets. This shows the known feature that the gluon jets have on average a softer fragmentation than quark jets at the same parton $p_{\mathrm{t}}$.

\section{\textit{NT90} and the UM}
\begin{figure}[htb]
\begin{center}
\includegraphics[width=17pc]{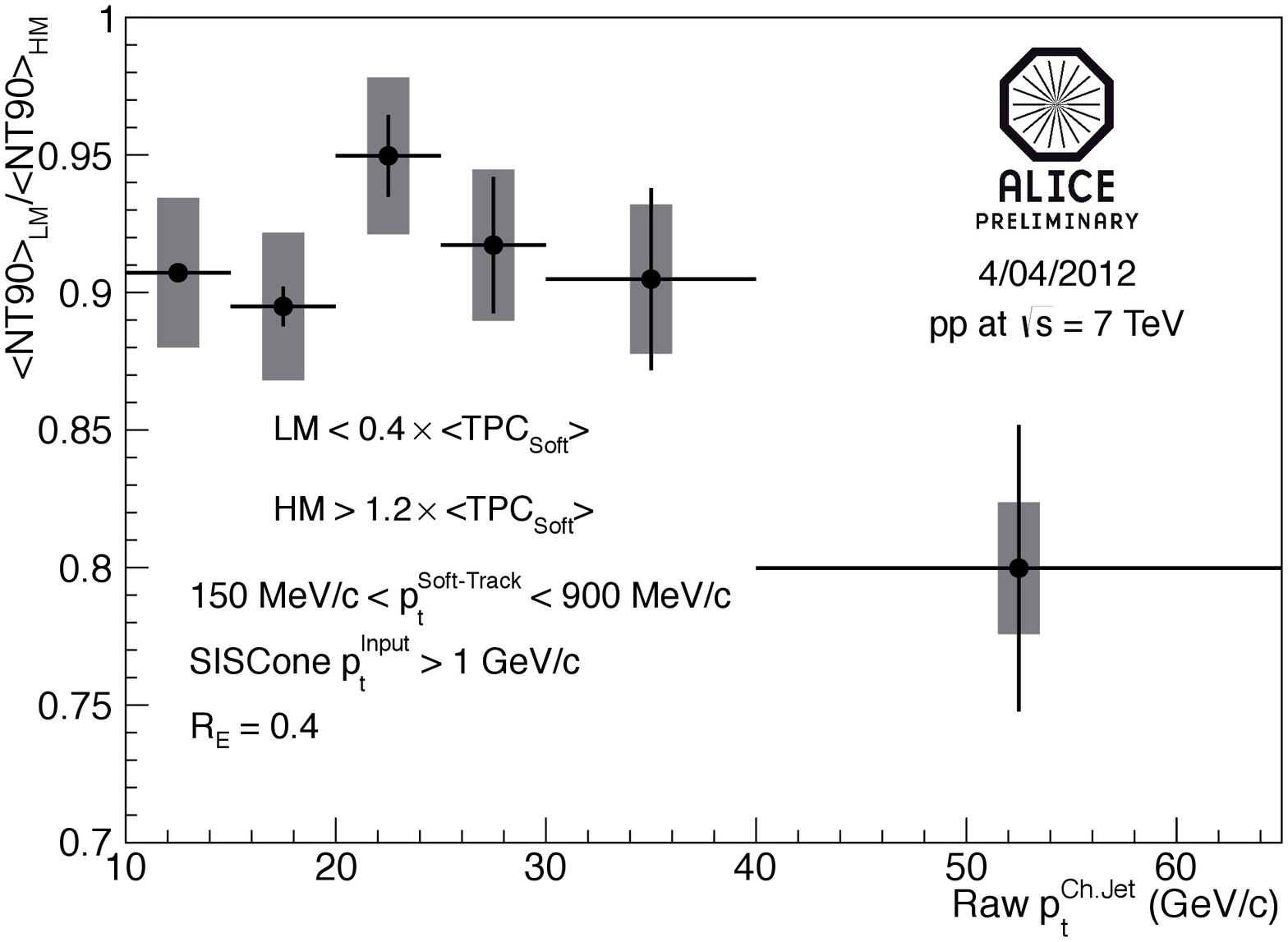}
\end{center}
\caption{\label{fig:figNT90pTData} Ratio of the mean values of \textit{NT90} obtained from two jet samples associate to different values of the UM as a function of the raw $p_{\mathrm{t}}^{\mathrm{Ch.Jet}}$. One jet sample is associated to values of the UM smaller than the average soft TPC multiplicity (LM) while the other is from jets associated to values of the UM larger than the average soft TPC multiplicity (HM). The data is presented at the detector level.}
\end{figure}

Figure \ref{fig:figNT90pTData} shows the ratio of the mean values of \textit{NT90} from jet samples corresponding to two bins of the UM as a function of the raw $p_{\mathrm{t}}^{\mathrm{Ch.Jet}}$. The ratio is between a bin with low UM, $\langle$\textit{NT90}$\rangle_{\mathrm{LM}}$, and a bin with high UM, $\langle$\textit{NT90}$\rangle_{\mathrm{HM}}$.  The bin with low UM has a soft TPC multiplicity smaller than 0.4 times the mean soft TPC multiplicity\footnote{The mean soft TPC multiplicity is the average number of tracks with 150 MeV/$c$ $< p_{\mathrm{t}}^{\mathrm{Soft-Track}} <$ 900 MeV/$c$ that are outside the areas related with jet fragmentation.} and the one with high UM is associated to a soft TPC multiplicity larger than 1.2 times the mean soft TPC multiplicity. From the figure one can see that the jet properties from the two jet samples are different. The mean values of the \textit{NT90} from the jets associated to high UM are larger than those associated to low values of the UM. This means that the jets in events with large UM have a softer fragmentation compared to the jets associated to low values of the UM.

\begin{figure}[!htb]
\begin{center}
$\begin{array}{c@{\hspace{0.5cm}}c}
\includegraphics[width=17pc]{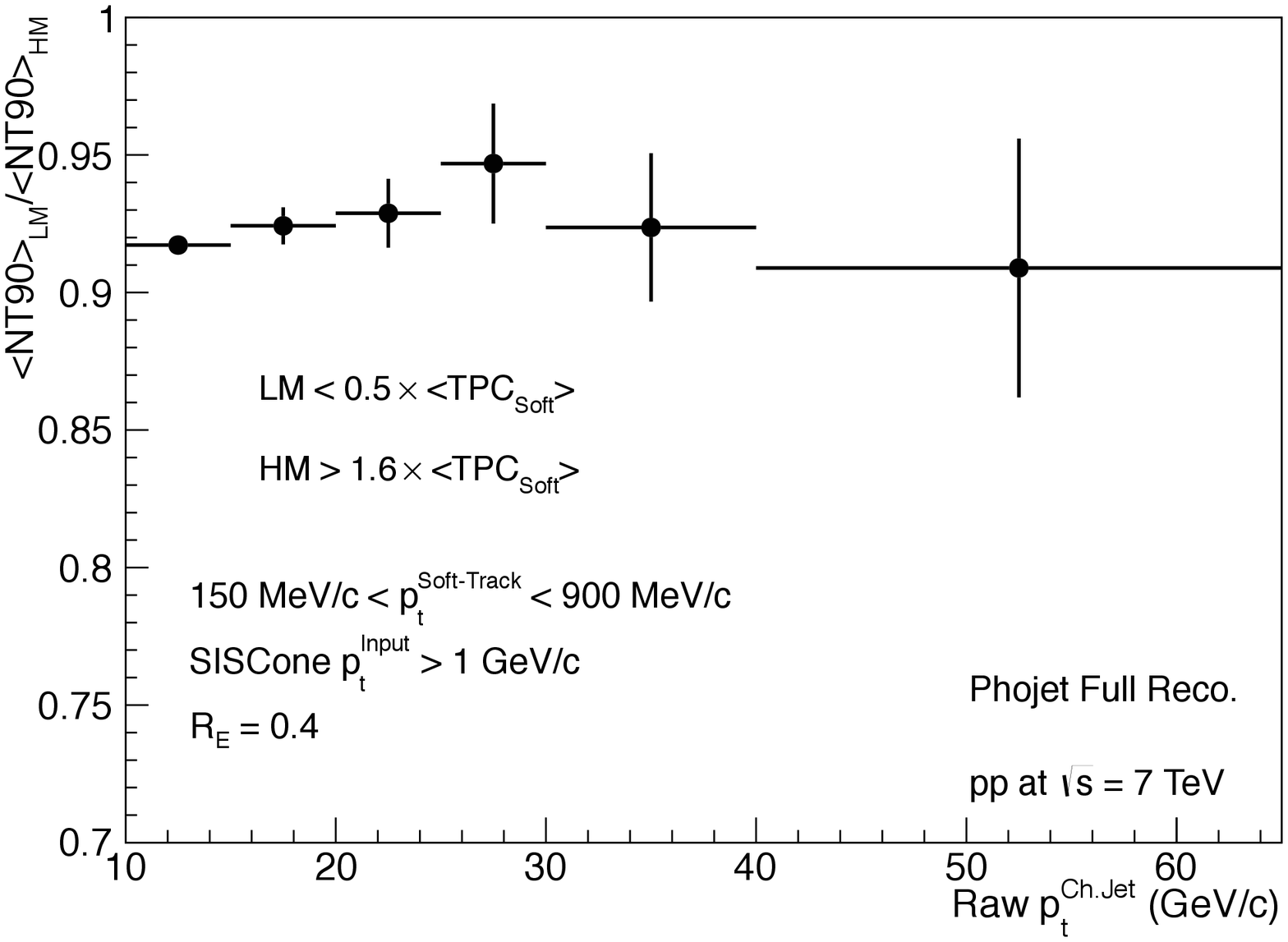} &
	\includegraphics[width=17pc]{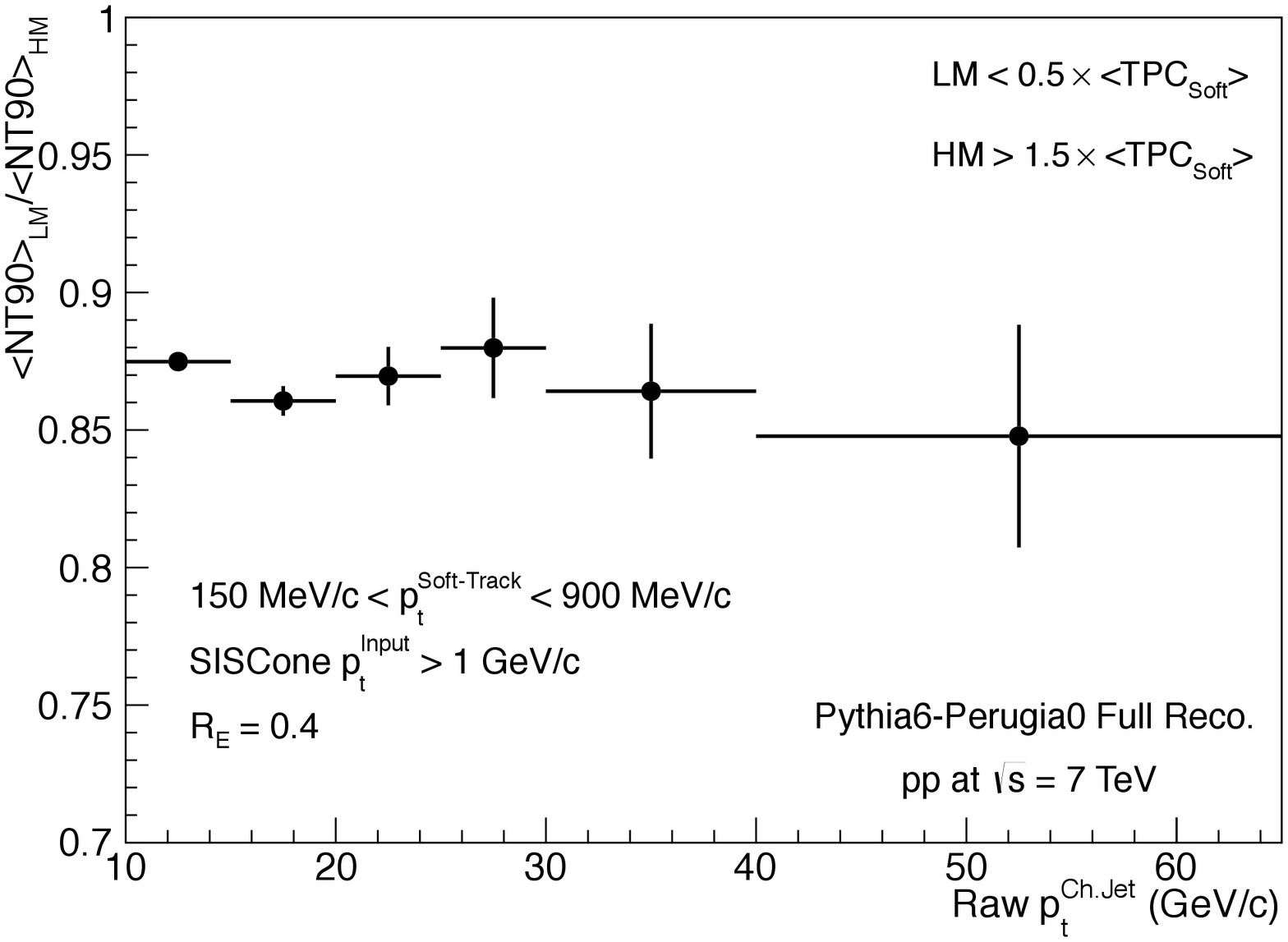} \\ [0.4cm]
\end{array}$
\end{center}
\caption{\label{fig:figNT90-FunctionJetpT-MC} Ratio of the mean values of \textit{NT90} from two bins of the UM as a function of the raw $p_{\mathrm{t}}^{\mathrm{Ch.Jet}}$. Left panel: Results obtained with a Phojet simulation including full detector simulation. Right panel: Results obtained with a Pythia6 simulation with the Perugia 0 tune including full detector simulation.}
\end{figure}

After looking at the differences in the properties of the charged jets from the two bins of UM, figure \ref{fig:figNT90-FunctionJetpT-MC} shows the results corresponding to those from figure \ref{fig:figNT90pTData} but now obtained from MC generators including full detector simulation. The left panel of figure \ref{fig:figNT90-FunctionJetpT-MC} shows the results obtained with Phojet \cite{Phojet}, and the right panel shows the results obtained with a Pythia6 \cite{Pythia6} simulation using the tune Perugia 0 \cite{Perugia0}. As one can see from the figure the MC generators are able to reproduce a general trend observed in the data, i.e. a softer fragmentation of the jets associated to high values of the UM compared to those associated to low UM.

\begin{figure}[!htb]
\begin{center}
$\begin{array}{c@{\hspace{0.5cm}}c}
\includegraphics[width=17pc]{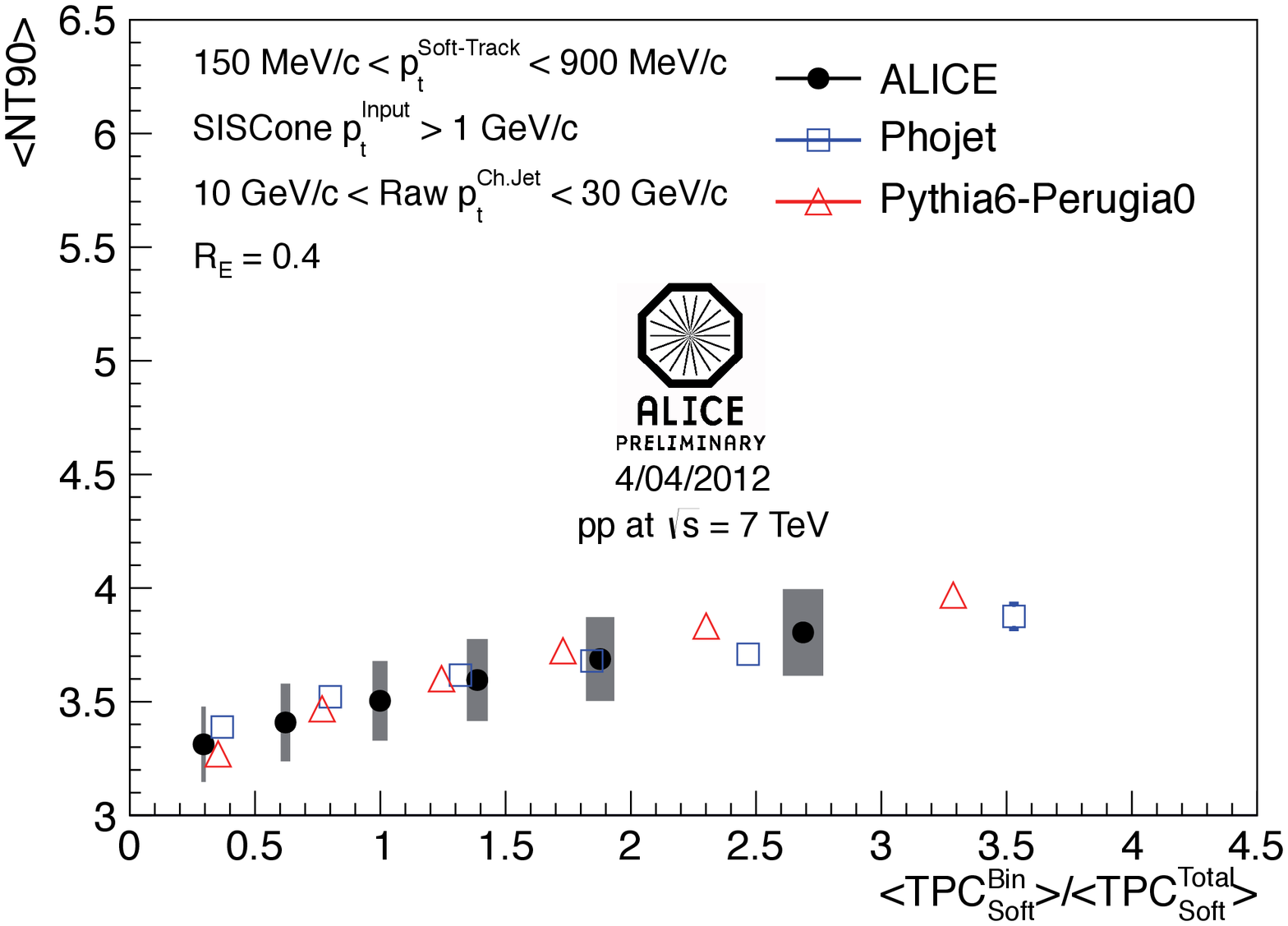} &
	\includegraphics[width=17pc]{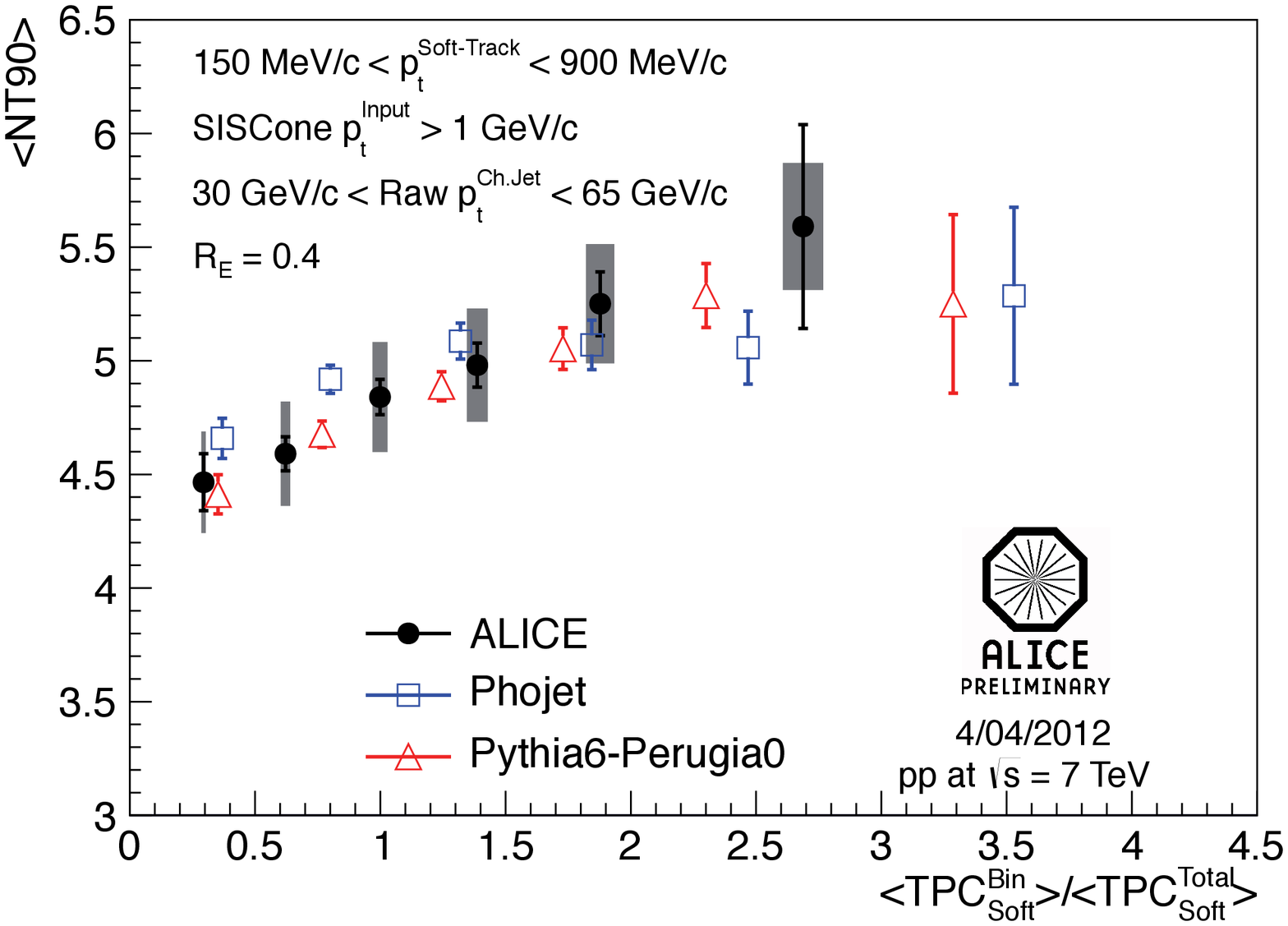} \\ [0.4cm]
\end{array}$
\end{center}
\caption{\label{fig:figNT90-Mult-DataMC} Mean values of \textit{NT90} as a function of the normalised UM. Left panel: Results obtained for jets with 10 GeV/$c$ $< p_{\mathrm{t}}^{\mathrm{Ch.Jet}} <$ 30 GeV/$c$. Right panel: Results for jets with 30 GeV/$c$ $< p_{\mathrm{t}}^{\mathrm{Ch.Jet}} <$ 65 GeV/$c$.}
\end{figure}

A comparison of the experimental results and the predictions from the MC generators is presented in figure \ref{fig:figNT90-Mult-DataMC}, which shows the mean values of \textit{NT90} as a function of the normalized UM measured in the ALICE central barrel. 

The standard way to present results as a function of multiplicity is by plotting the observable as a function of fractions of the mean multiplicity. In this case the results are presented for different bins of UM, $\langle$ TPC$_{\mathrm{Soft}}^{\mathrm{Bin}}$ $\rangle$, normalized by the mean value of UM, $\langle$ TPC$_{\mathrm{Soft}}^{\mathrm{Total}}$ $\rangle$, measured in the central barrel for each data sample. The data was separated into two ranges of $p_{\mathrm{t}}^{\mathrm{Ch.Jet}}$. The left panel of figure \ref{fig:figNT90-Mult-DataMC} shows the results obtained for 10 GeV/$c$ $< p_{\mathrm{t}}^{\mathrm{Ch.Jet}} <$ 30 GeV/$c$ and in the right panel for jets with 30 GeV/$c$ $< p_{\mathrm{t}}^{\mathrm{Ch.Jet}} <$ 65 GeV/$c$. The results obtained from the MC generators are presented after the full ALICE detector simulation so they are directly comparable with the experimental results. For jets with 10 GeV/$c$ $< p_{\mathrm{t}}^{\mathrm{Ch.Jet}} <$ 30 GeV/$c$ Pythia6 shows a similar dependence with the multiplicity as the one observed in the experimental data, while Phojet shows a flatter dependency. For jets with 30 GeV/$c$ $< p_{\mathrm{t}}^{\mathrm{Ch.Jet}} <$ 65 GeV/$c$ one can observe a monotonic increase of the mean values of \textit{NT90} with increasing multiplicity for the experimental data. A similar trend is observed in the Pythia6 results while Phojet on the other hand shows an increase of the mean values of \textit{NT90} that seems to saturate after reaching the mean value of the UM.

\section{Conclusions}

We have presented an estimator of the underlying soft multiplicity in pp collisions. The characterization is done with the UM, that uses the information from a jet finder to localize the regions of the detector acceptance that contain most of the jet fragmentation products. Then we separated the collision data in bins of UM and studied the charged jet fragmentation using the \textit{NT90} observable. We observe that events with high activity of the UM are correlated to jets that undergo a softer fragmentation. The MC models Pythia6 and Phojet are able to reproduce the experimental observations for low momentum jets, while for intermediate jet transverse momentum Phojet seems to fail for UM above the mean values. The fact that the MC generators are able to describe the dependence of the jet fragmentation as a function of the UM needs further studies given that the models contain a simplified approach to describe the nucleon-nucleon collision centrality.


\section*{Acknowledgments}

The author thanks C. Blume for the collaboration in the analysis presented in this work and C. Klein-B\"osing and M. van Leeuwen for their comments and support. This work was supported by the Helmholtz Research School for Quark Matter Studies.


\smallskip

\end{document}